\documentclass[12pt]{article}

\usepackage{amsmath,amssymb,amsfonts,cite}
\usepackage{graphicx,epsfig}
\usepackage{color}
\usepackage{amsmath}
\usepackage{amsfonts}
\usepackage{amssymb}
\usepackage{float}
\usepackage{cancel}
\usepackage{caption}
\usepackage{graphicx}

\newcommand{\beq}{\begin{eqnarray}}
\newcommand{\eeq}{\end{eqnarray}}

\newcommand{\centeron}[2]{{\setbox0=\hbox{#1}\setbox1=\hbox{#2}\ifdim
\wd1>\wd0\kern.5\wd1\kern-.5\wd0\fi \copy0
\kern-.5\wd0\kern-.5\wd1\copy1\ifdim\wd0>\wd1
                                  \kern.5\wd0\kern-.5\wd1\fi}}
\newcommand{\ltap}{\>\centeron{\raise.35ex\hbox{$<$}}
                          {\lower.65ex\hbox{$\sim$}}\>}
\newcommand{\gtap}{\>\centeron{\raise.35ex\hbox{$>$}}
                          {\lower.65ex\hbox{$\sim$}}\>}

\newcommand\ZZ{\hbox{\zfont Z\kern-.4emZ}}
\font\zfont = cmss10 

\newcommand{\fref}[1]{Fig.\ \ref{f.#1}}
\newcommand{\eref}[1]{eq.\ (\ref{e.#1})}

\newcommand{\cref}[1]{Chapter \ref{c.#1}}

\def\nn{\nonumber \\}

\def\beq{\begin{equation}}
\def\eeq{\end{equation}}
\newcommand{\ba}{\begin{array}}
\newcommand{\ea}{\end{array}}
\newcommand{\bea}{\begin{eqnarray}}
\newcommand{\eea}{\end{eqnarray} }
\newcommand{\bal}{\begin{align}}
\newcommand{\eal}{\end{align}}
\def\bi{\begin{itemize}}
\def\ei{\end{itemize}}
\def\ben{\begin{enumerate}}
\def\een{\end{enumerate}}
\def\beq{\begin{equation}}
\def\eeq{\end{equation}}
\def\bc{\begin{center}}
\def\ec{\end{center}}
\def\bt{\begin{table}}
\def\et{\end{table}}
\def\btb{\begin{tabular}}
\def\etb{\end{tabular}}
\newcommand{\bvec}{\left ( \ba{c}}
\newcommand{\evec}{\ea \right )}

\def\cl{{\mathcal L}}

\def\co{{\mathcal O}}

\def\cH{{\mathcal H}}

\def\gev{\, {\rm GeV}}

\def\mass2{mass${}^2$}

\def\msusy{M_{\rm soft}}

\def\ra{\rangle}
\def\la{\langle}

\def\pa{\partial}

\def\simlt{\stackrel{<}{{}_\sim}}
\def\simgt{\stackrel{>}{{}_\sim}}

\newcommand{\ti}{\tilde}

\def\ov{\overline}

\def\eps{\epsilon}

\begin{document}
\begin{titlepage}
\vskip1.5cm
\vskip1.cm
\begin{center}
{\huge \bf Buried Higgs}
\vspace*{0.1cm}
\end{center}
\vskip0.2cm

\begin{center}
{\bf Brando Bellazzini$^{a}$, Csaba Cs\'aki$^{a}$, Adam Falkowski$^{b}$, and Andreas
Weiler$^{c}$}

\end{center}
\vskip 8pt

\begin{center}
$^{a}$ {\it Institute for High Energy Phenomenology\\
Newman Laboratory of Elementary Particle Physics\\
Cornell University, Ithaca, NY 14853, USA } \\
\vspace*{0.3cm}
$^b$ {\it NHETC and Department of Physics and Astronomy \\
Rutgers University, Piscataway, NJ 088550849, USA } \\
\vspace*{0.3cm}
$^c$ { \it CERN Theory Division, CH-1211 Geneva 23, Switzerland} \\

\vspace*{0.3cm}

{\tt b.bellazzini@cornell.edu,  csaki@cornell.edu, falkowski@physics.rutgers.edu, andreas.weiler@cern.ch}
\end{center}

\vglue 0.3truecm

\begin{abstract}
\vskip 3pt \noindent
We present an extension of the MSSM where the dominant decay channel of the Higgs boson is a cascade decay into a four-gluon final state.
In this model the Higgs is a pseudo-Goldstone boson of a broken global symmetry $SU(3)\to SU(2)$.
Both the global symmetry breaking and electroweak symmetry breaking are radiatively induced.
The global symmetry breaking pattern also implies the existence of a light (few GeV) pseudo-Goldstone boson $\eta$ which is a singlet under the standard model gauge group. The $h \to \eta \eta$ branching fraction is large, and typically dominates over the standard $h \to b b$ decay. The dominant decay of $\eta$ is into two gluons, while the decays to photons, taus or lighter standard model flavors are suppressed at the level of $10^{-4}$ or more.
With $h\to$ 4 jets as the dominant decay, the Higgs could be as light as 78 GeV without being detected at LEP, while detection at the LHC is extremely challenging.
However many of the super- and global symmetry partners of the standard model particles should be easily observable at the LHC.
Furthermore, the LHC should be able to observe a ``wrong" Higgs that is a 300-400 GeV heavy Higgs-like particle with suppressed couplings to W and Z that by itself does not account for electroweak precision observables and the unitarity of WW scattering.
At the same time, the true Higgs is deeply buried in the QCD background.

\end{abstract}

\end{titlepage}

\subsection*{Introduction}

The elegant and appealing idea of low-energy supersymmetry faces technical difficulties when confronted with experimental data.
In the MSSM at tree level the mass of the Higgs boson is bounded from above by the Z boson mass.
In order to reconcile the MSSM with the non-discovery of the Higgs at LEP one has to assume that supersymmetry breaking introduces large loop corrections to the Higgs quartic self-coupling (which sets the Higgs boson mass).
However, the same large parameters that lift the Higgs boson mass also contribute to the W and Z mass (that is to the Higgs VEV), and large accidental cancellations are required to keep the Z-mass at its experimental value.
In a typical scenario this leads to a fine-tuning at the level of 1\% or worse.
Additional theoretical structures beyond those of the MSSM are necessary if supersymmetry is to provide a solution to the naturalness problem.
The most popular approach is to engineer new tree-level contributions to the quartic Higgs potential, which is possible e.g in singlet (NMSSM)~\cite{NMSSM} or gauge~\cite{dterms}  extensions of the MSSM.

Another interesting possibility for improving the naturalness of low-energy supersymmetry is to assume that
the Higgs is a pseudo-Goldstone boson (pGB) of a spontaneously broken approximate global symmetry, much as in Little Higgs models \cite{LH}, but now in a supersymmetric context. This approach~\cite{BCG,BCFP,RS,CMSS,FPS,BPRV,BCDW} is referred to as the {\em double protection} of the Higgs potential or as the {\em super-little Higgs}\footnote{For early attempts of supersymmetric theories with a pGB Higgs see \cite{Dvali1,Dvali2}}.
The collective symmetry breaking pattern of little Higgs models combined with supersymmetry indeed implies that one-loop corrections to the Higgs mass are completely finite, thereby greatly reducing one-loop corrections to the W and Z mass, and improving on naturalness. Unfortunately, due to the reduced sensitivity of the Higgs potential to loop effects it is quite difficult to obtain a Higgs heavier than 114.4 GeV in these models. Thus, additional (sometimes truly baroque)  structures have to be invoked to make the Higgs heavy enough, with the simplest complete model involving a U(1) gauge extension~\cite{BCDW}.

However the Higgs does not necessarily have to be as heavy as 115 GeV if it decays in a non-standard way.
Dermisek and Gunion \cite{DG1} (see also \cite{DM} and \cite{Carena:2002bb}) 
pointed out that the LEP bounds can be relaxed if the Higgs undergoes a cascade decay to a many particle final state, rather than directly decaying into a pair of SM particles.
This is possible if for example the Higgs can first decay to a pair of light singlet pseudoscalars, each of which subsequently decays to a pair of quarks or leptons.
Depending on the dominant decay channels of the pseudoscalar the LEP bound on the Higgs mass can be as low as 86 GeV for a $4\tau$ final state~\cite{LEP4b}, or almost as high as the standard bound 110 GeV for $4b$ final states~\cite{LEP4b} (assuming the standard model (SM) production cross section for the Higgs), see \cite{CDGW} for a review.
The cascade decays of the Higgs are most often realized in the context of the NMSSM \cite{NMSSM2}, but there exist also non-supersymmetric realizations \cite{GPRS,Cheung:2006nk}.
A Higgs boson mass below 100 GeV  would greatly reduce the number of problems that are plaguing model building and could eradicate the little hierarchy problem.

\begin{figure}[tb]
\bc
\includegraphics[width=0.38\textwidth]{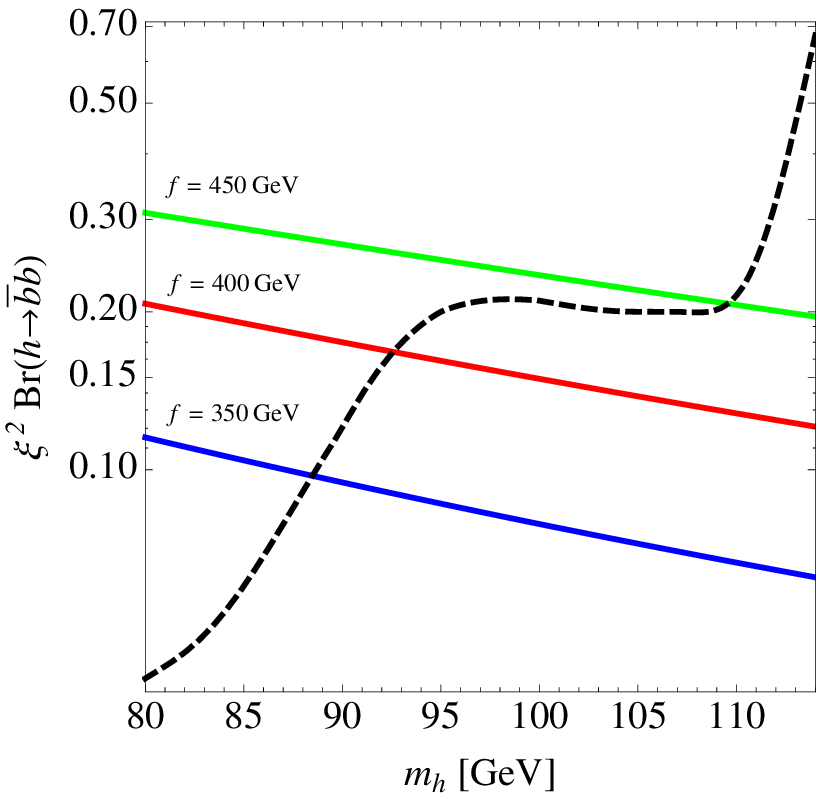} \hspace*{1.cm}
\includegraphics[width=0.4\textwidth]{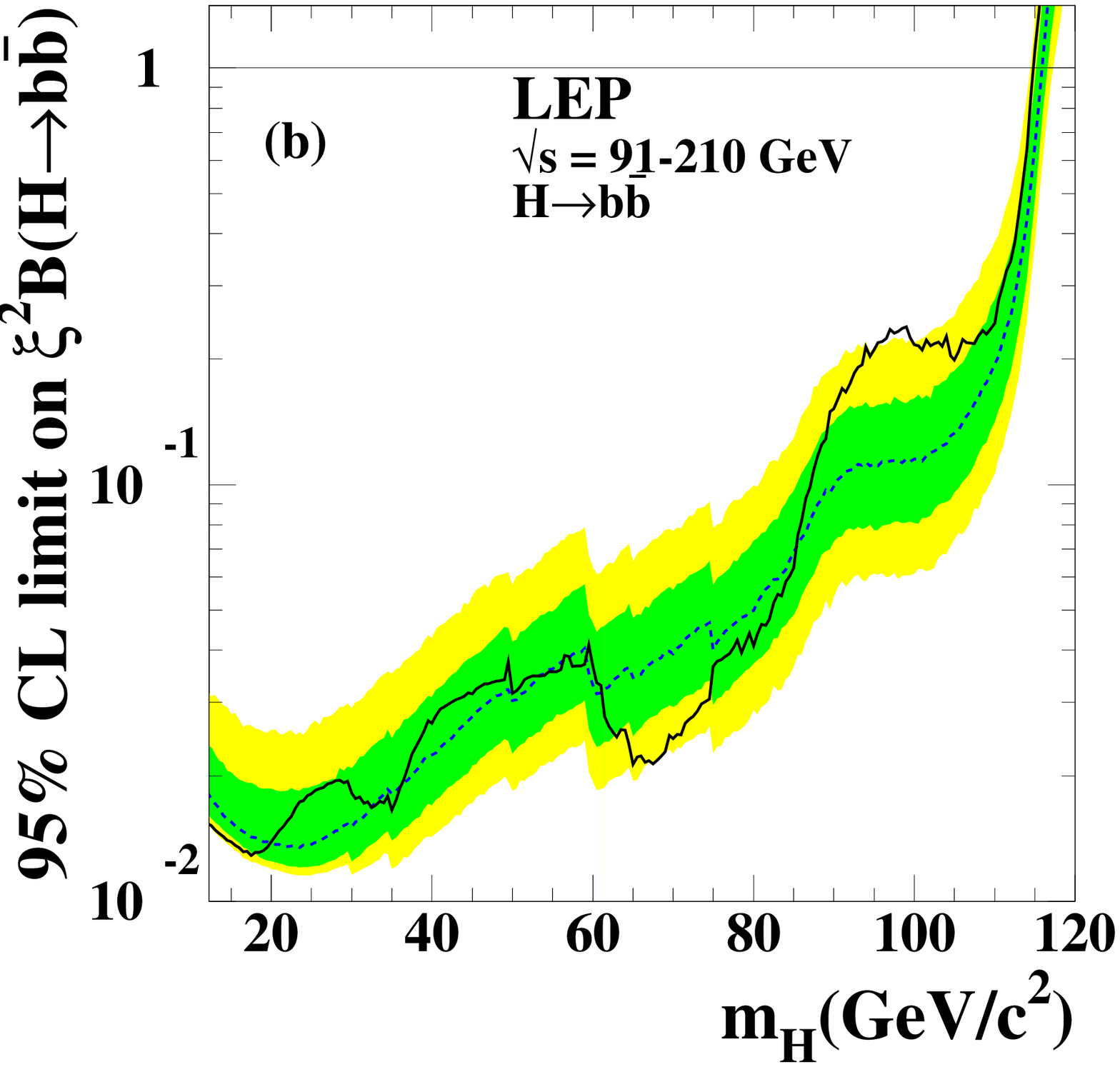}
\ec
\caption{The parameter $\xi^2 {\mathrm{BR}}(h \to b \ov b)$ in this model for 3 representative values of the global symmetry breaking scale
as a function of the Higgs mass.
The dashed line is an approximation of the observed LEP bound transcribed from the actual LEP plot reprinted from \cite{LEP4b} on the right hand side. We can see that while for $f=450$ GeV the bound is over 110 GeV for the Higgs mass, for $f=350$ GeV the Higgs could be lighter than 90 GeV.
\label{f.brbb}}
\end{figure}

The main point of this paper is to argue that the idea of Dermisek and Gunion is very naturally realized in the double protection models: all the necessary ingredients are already in there.
Since the simplest models of that type are based on the $SU(3) \to SU(2)$ pattern of spontaneous global symmetry breaking one ends up with five pGBs.
Four of these are identified with the Higgs doublet, while the remaining one (which is henceforth referred to as $\eta$) is a singlet under the SM gauge interactions, and naturally has the properties of the Dermisek-Gunion pseudoscalar needed to hide the Higgs at LEP!
The singlet $\eta$ receives a mass from one-loop corrections which naturally fall into the few GeV range, thus allowing the $h\to 2 \eta$ decay.
Furthermore, a pair of the pGB singlets has a higher-derivative tree-level interaction vertex with the Higgs boson that is suppressed by the global symmetry breaking scale $f$, which allows the Higgs cascade decays to have a large branching fraction. For typical parameters of the model when the scale $f$ is not much larger than the electroweak scale, the branching ratio for the two-body Higgs decays to the SM fermions is less than 20\% (see Fig.\ref{f.brbb}).
This is enough to hide an 80-100 GeV Higgs from LEP\footnote{It may also provide the explanation of the large excess of Higgs like events  seen at LEP for a Higgs mass of about $98$ GeV \cite{Dermisek:2005gg}.} provided the pGB singlet is lighter than 9.2 GeV and therefore cannot decay to b-quarks.
To calculate the $\eta$ mass one needs to specify the fermion structure of the theory.
We show that with the same matter content as \cite{BCDW},  a large fraction of the $\eta$ masses lies naturally in the interesting range below 9.2 GeV. The actual Higgs phenomenology is then determined by the leading decays of $\eta$.
We argue that in our model the gluon decay channel is by far the dominant one, resulting in the $h\to 4j$ signal that is in fact invisible at LHC due to the huge background.
Thus, double protection predicts very peculiar experimental signatures where a host  of super- and little partners are visible at the LHC but the Higgs boson responsible for electroweak breaking remains elusive.
However, a ``wrong" Higgs may show up more easily at the LHC: the theory predicts a heavy (300-400 GeV) scalar particle, corresponding to the oscillations of the global symmetry breaking scale $f$, whose production cross section is 15-25\% that of the SM Higgs boson. 
This {\em radial mode} is similar to the Higgs in many respects, but it has reduced couplings to W and Z gauge bosons, thus being unable on its own to account for the electroweak precision fits and unitarization of gauge boson scattering amplitudes.

The paper is organized as follows:
first we introduce the gauge and global symmetry breaking structure and identify the Goldstones in the theory. We then calculate the $h\to 2\eta$ branching fraction and compare it to the leading SM $h\to 2b$ channel.
We specify the Yukawa structure of the theory and we calculate the $\eta$ mass.
We present the distribution of the $\eta$ and Higgs masses and the necessary fine tuning needed to achieve those values.
Then we show that  $\eta\to gg$ is the dominant decay channel, resulting in the Higgs decaying into 4 jets.
Finally we discuss the impact of this model on the electroweak precision observables and on unitarity, and then conclude.

\subsection*{Gauge and Global Symmetries, Goldstones}

The model we consider is based on the $SU(3)_C \times SU(3)_W \times U(1)_X$ gauge symmetry \cite{BCFP,RS,CMSS,BPRV,BCDW} which is the supersymmetric version of the simplest little Higgs model of~\cite{Martin}.
$SU(3)_W \times U(1)$  is broken by two vectorlike sets  $\Phi_{u,d}$ and $\cH_{u,d}$ of Higgs superfields with the following quantum numbers
\beq
\label{e.pgb}
\begin{array}{c|c|c|c|}
& SU(3)_C & SU(3)_W & U(1)_X
\\
\hline
\cH_u,\Phi_u  & 1 & 3      &  1/3
\\
\cH_d,\Phi_d  & 1 & \bar 3 & -1/3
\end{array}
\eeq
The central assumption here is that there are no cross-couplings between the two sets of Higgses,
that is the mass terms $\Phi_d \cH_u$ and  $\cH_d \Phi_u$ in the superpotential are absent or very suppressed even though they are allowed by the gauge symmetry.
It should be stressed that this assumption is technically natural in the supersymmetric context thanks to the non-renormalization theorems.
The consequence of this assumption is that there is an approximate $SU(3)_1 \times SU(3)_2$ global symmetry (broken to the diagonal $SU(3)_W$ by the gauge interactions) acting separately on the two Higgs sets.
$\Phi_{u,d}$ are assumed to have a large VEV (generated by some additional supersymmetry preserving dynamics) in the 10 TeV regime,
\beq
\la \Phi_u \ra^T =  \la \Phi_d \ra = (0,0,F/\sqrt{2}) .
\eeq
This VEV determines the orientation of the SM group within the SU(3)$\times$U(1)$_X$ and
breaks the gauge group down to $SU(3)_C \times SU(2)_W \times U(1)_Y$. In our convention $SU(2)_W$ is acting on the upper two components of the $SU(3)_W$ triplets and hypercharge is realized as  $Y =  T_8/\sqrt{3} + X$ with $T_8=\frac{1}{2\sqrt{3}} {\rm diag} (1,1,-2)$.
The VEV also breaks the $SU(3)_1$ global symmetry with all the Goldstone bosons being eaten by the $SU(3)\times U(1)/SU(2)\times U(1)$ massive gauge fields, but it leaves  $SU(3)_2$ intact.
The other set of Higgs triplets $\mathcal{H}_{u,d}$ is assumed to get much smaller VEVs in the $f=300-500$ GeV range (generated radiatively, much like the Higgs VEV in the MSSM). This VEV will also break the SU(3)$\times$U(1) gauge symmetry to SU(2)$\times$U(1) and produce its own 5 Goldstone bosons. If the two sets of VEVs are somewhat misaligned then the only remaining unbroken gauge group is U(1)$_{em}$, with the misalignment responsible for electroweak breaking. 
In the limit $F\gg f$ the misalignment between the two sets of VEVs is parameterized by the 5 Goldstone bosons as
\begin{equation}
\mathcal{H}_u= e^{i\Pi /f} f \sin\beta \left( \begin{array}{c} 0 \\ 0 \\ 1 \end{array} \right) , \ \ \mathcal{H}_d= e^{-i\Pi/f} f \cos\beta \left( \begin{array}{c} 0 \\ 0 \\ 1 \end{array} \right) ,
\end{equation}
where the pion matrix corresponds to the broken generators
\begin{equation}
\Pi = \left( \begin{array}{ccc|c} & & & \\ & & & H\\ & & & \\ \hline & H^\dagger & &  \frac{\tilde{\eta}}{\sqrt{2}} \end{array} \right) .
\end{equation}
3 of the 5 Goldstones are eaten by the W and Z bosons after electroweak symmetry breaking, with two real physical pGBs $\tilde{h}, \tilde{\eta}$ remaining in the physical spectrum (the tilde here is to stress that the field is not canonically normalized; the properly normalized fields $h$, $\eta$ will be defined below). In terms of the physical Goldstones the parametrization of the triplets is given by\footnote{Taking into account $1/F$ corrections there is also a small component of the physical pGBs embedded in $\Phi_{u,d}$.}
\beq
\label{e.pgbe}
\cH_u = f \sin\beta \bvec 0 \\ \sin(\ti h/\sqrt {2}f) \\ e^{i \ti \eta/\sqrt {2}f} \cos(\ti h/\sqrt {2}f) \evec
\qquad
\cH_d^T = f \cos\beta  \bvec 0 \\ \sin(\ti h/\sqrt {2}f) \\ e^{-i \ti \eta/\sqrt {2}f} \cos(\ti h/\sqrt {2}f) \evec .
\eeq
The real field $\ti h$ is the pGB Higgs boson whose VEV breaks the electroweak symmetry.
The electroweak scale $v_{EW} = 174$ GeV is related to the Higgs VEV $\la \ti h \ra  = \sqrt{2} \ti v$  by
\begin{equation}
v_{EW} = f \sin(\ti v/f) \,.
\end{equation}
The other pGB field $\eta$ lives fully in the third component of the triplet, therefore it is a perfect singlet under the SM gauge interactions. Thus there are no constraints on $\eta$ from a contribution to the Z-width.\footnote{Once $\eta$ gets a mass there will be a small mixing with the physical pseudoscalar $A$ living in the Higgs doublets, but the mixing angle is suppressed by $m_\eta^2/m_A^2 \simlt 10^{-5}$.}

\subsection*{Higgs decays: $h\to 2 \eta$ vs. $h\to b\bar{b}$}

Next we discuss the Higgs decay modes, and argue that there is a possibility for the $h\to \eta \eta$ mode to dominate for generic values of the parameters.
Even though  $\eta$ is an SU(2) singlet, it does have a tree-level derivative coupling to the Higgs field $h$ due to $h$ partly living in the third component of $\mathcal{H}_{u,d}$ (and {\it not} because of $\eta$ mixing into the doublet part of $\mathcal{H}_{u,d}$). The symmetry preserving derivative coupling (characteristic to exact Goldstone bosons) originates from the Higgs kinetic terms,
\beq
\cl_{pGB} \approx {1 \over 2}(\pa_\mu \ti h)^2 + {1 \over 2} \cos^2(\ti h/\sqrt{2} f) (\pa_\mu \ti \eta)^2.
\eeq
At one loop there are also non-derivative interactions via the Coleman-Weinberg potential that depend on both $h$ and $\eta$, but these lead to subleading interaction terms.
After the Higgs gets a VEV we define the canonically normalized Higgs boson field $h$ and the singlet field $\eta$ by $\ti h = \sqrt{2} \ti v + h$,  $\ti \eta = \eta/\cos(\ti v/f)$.
This leads to the following vertex of the Higgs boson with two singlets:
\beq
\cl_{h\eta^2} \approx - h (\pa_\mu \eta)^2  {\tan(\ti v/f) \over \sqrt{2} f}.
\eeq
The decay width of the Higgs boson into two singlets is given by
\beq
\Gamma_{h \to \eta\eta} \approx {1 \over 64 \pi} \left( 1 - {v_{EW}^2 \over f^2} \right)^{-1}
{m_h^3 v_{EW}^2\over f^4}.
\eeq
The coupling of the Higgs boson  to the SM quarks and leptons is the same as in  the SM, up to an additional factor $\cos (\ti v/f)$ that arises due to its pGB nature.
Thus, the decay width into a pair of SM fermions is given by
\beq
\Gamma_{h \to f \ov f}
=  \left( 1 - {v_{EW}^2 \over f^2} \right)\Gamma_{h \to f \ov f}^{SM}
= c_{QCD} {N_c \over 16 \pi}  \left( 1 - {v_{EW}^2 \over f^2} \right) {m_h m_f^2 \over v_{EW}^2}. 
\eeq
Here, $N_c = 3$ for quarks and $1$ for leptons.
$c_{QCD}$ arises due to  higher order QCD corrections which  can be numerically important; for example for the b-quark it is given by $c_{QCD} \approx 1/2$ for a 100 GeV Higgs \cite{DKS}.
The relevant quantity for LEP searches, customarily denoted as $\xi^2 \mathrm{BR}(h\to b \ov b)$,  is the branching ratio for a decay into $b$ quarks multiplied by the suppression of the Higgs production cross section.
The latter should also be taken into account in  our model because the coupling of the pGB Higgs to the Z boson is suppressed, much as the Higgs-fermion coupling, by a factor  $\cos(\ti v/f)$.
It then follows
\beq
\xi^2 {\mathrm{BR}}(h \to b \ov b) \equiv
{\sigma(e^+ e^- \to Z h ) \over \sigma_{SM}(e^+ e^- \to Z h )} {\mathrm{BR}}(h \to b \ov b)
= {\Gamma^{SM}_{h \to b \ov b} \over \Gamma_{h \to \eta\eta}
+ \left(1 - {v_{EW}^2 \over f^2}\right) \sum_f \Gamma^{SM}_{h \to f \ov f}}
\left(1 - {v_{EW}^2 \over f^2}\right)^2
\eeq
We plot $\xi^2 {\mathrm{BR}}(h \to b \ov b)$ as a function of the Higgs mass for several choices of the global symmetry breaking scale $f$, together with the combined LEP bound on $\xi^2 {\mathrm{BR}}(h \to b \ov b)$ from~\cite{LEP4b}.
If $f$ is as small as 350 GeV, the $b \ov b$ branching ratio is sufficiently suppressed to allow for a Higgs as light as the Z boson.
Once $f$ is raised to around $450$ GeV or higher, the generic 114.4 GeV limit from LEP cannot be significantly relaxed - the $b \ov b$ branching ratio becomes large enough to have been observable at LEP.

\subsection*{Matter Yukawas}

Of course, in order to make the Higgs decay into $\eta$ one has to ensure that the latter is light enough. Thus we now turn to discussing how the $\eta$  mass is generated.
At tree level the $\eta$ mass as well as the masses of the remaining four pGBs vanish, but they are generated at one loop.
The leading contributions are expected from the third generation quarks and their symmetry partners (superpartners and global symmetry partners).
Following \cite{BCDW}, we consider the following embedding of the third generation quarks and leptons into $SU(3)_C \times   SU(3)_W \times U(1)_X$ representations
\beq
\begin{array}{l|c|c|c|}
& SU(3)_C &  SU(3)_W & U(1)_X
\\
\hline
Q = (t^Q,b^Q,\hat b^Q)& 3 & 3 & 0
\\
V = (b^V,t^V,\hat t^V) & 3 & \bar 3 & 1/3
\\
V_c = (b_c^V,t_c^V,\hat t_c^V) & \bar 3 & 3 & -1/3
\\
t_c & \bar 3 & 1 & -2/3
\\
b_{c}^{1,2} & \bar 3 & 1 & 1/3
\\
\hline
L_{1,2} = (\tau_{1,2}^L,\nu_{1,2}^L,\hat \nu_{1,2}^L) &  1 & \bar 3 & - 1/3
\\
E_c =   (\nu_c^E,\tau_c^E,\hat \tau_c^E )&  1 & \bar 3 & 2/3
\\
\nu_{c}^{1,2,3} &  1 & 1 & 0
\end{array}
\eeq
This assignment of  representations is anomaly free.
The quark and lepton masses originate from the Yukawa couplings and the supersymmetric mass terms
\beq
\label{e.wtb}
W = y_1 t_c V \Phi_u  + y_2 \cH_u V_c Q  + \mu_V V V_c  + y_{b1} \Phi_d Q  b_c^1 + y_{b2} \cH_d Q b_c^2
+ y_{\tau 1} \Phi_d L_1 E_c + y_{\tau 2} \cH_d L_2 E_c  .
\eeq
More Yukawa and mass terms are needed to give masses to all neutrinos but we are not concerned with it here.
As in the case of the triplet Higgs superpotential, these are not the most general Yukawa couplings consistent with the gauge symmetries,
in particular  $\ti y_1 t_c V \Phi_u$ and  $\ti y_2 \cH_u V_c Q$ are omitted.
Omitting those and other allowed terms amounts to imposing a collective breaking of the {\em global} $SU(3)_2$ symmetry which acts on $\cH_{u,d}$ and remains after gauge symmetry breaking via $\Phi_{u,d}$ VEVs.
Note that in the top sector $SU(3)_2$ is restored  if any of the three couplings: $y_1$, $y_2$ or $\mu_V$ is set to zero.
At the same time, in the bottom sector  $SU(3)_2$ is restored if either $y_2$ or $y_{b1}$ vanishes.
The latter means that the bottom loops induce corrections of order $ y_2^2 y_{b1}^2 \log \frac{y_{b1} F}{M_{soft}^2}$ to the Higgs mass.
Since $F \sim 10$ TeV, an order one value of $y_{b1}$ starts reintroducing the little hierarchy problem due to the large $\log(y_{b1} F)$,
and in the following we assume $y_{b1} < 0.1$ to keep fine-tuning under control.
On the other hand, in the top sector all Yukawa couplings can be order one, as long as $\mu_V$ is less than TeV.

\subsection*{Radiative symmetry breaking and fine tuning}

The top Yukawa and mass terms included in \eref{wtb} at one-loop lead to radiative generation of the global symmetry breaking scale $f$ and the electroweak scale $v_{EW}$.
The former arises as a consequence of the negative contributions to the mass and quartic terms of the
triplet $\cH_u$,
\bea
m_{\cH_u}^2 &\approx&  - {3  y_2^2 \sin^2\beta \over 2 \pi^2} \msusy^2 \log(\Lambda/ M_{T}), 
\nn
\lambda_{\cH_u} &\approx& {3 y_2^4 \sin^4\beta  \over 8 \pi^2} \log((\msusy^2 + M_{T}^{2})/M_{T}^{2}),
\label{tripletpot}
\eea
where $M_T =\sqrt{\mu_V^2 + \sin^2\beta y_2^2 f^2}$ is the mass of the heavy fermionic top partner,
and $\msusy$ is the soft supersymmetry breaking scale (we assumed the common soft mass for all the stops and $F \gg f$).
The potential (\ref{tripletpot}) also generates the mass $m_r^2\sim 4 \lambda_{\mathcal{H}_u} f^2$ for the radial mode of the triplet $\mathcal{H}_u$ corresponding to the fluctuations of the VEV $f$.
This part of the potential is in many respects similar to generating the Higgs potential in the MSSM.
In particular, the mass term is logarithmically divergent and proportional to the soft supersymmetry breaking scale.
Yet it does not lead to the fine-tuning problem at the same level as in the MSSM.
This is because 1) the scale $f$ is larger than the electroweak scale 2)
we are free to take the Yukawa coupling $y_2$ to be larger than the SM top Yukawa coupling. One can define the amount of fine tuning necessary to maintain the hierarchy between $F$ and $f$ as the ratio of the physical radial mass to the loop induced correction of the triplet
\begin{equation}
FT_3= \frac{m_r^2/2}{|m^2_{\mathcal{H}_u}|} \sim \frac{y_2^2 f^2}{M_{soft}^2} \frac{ \log \frac{M_{soft}^2+M_T^2}{M_T^2}}{\log \frac{\Lambda^2}{M_T^2}}.
\end{equation}
For example, for $f \sim 350$ GeV and $y_2 \sim 1.8$ the fine-tuning is usually in the 5-10\% range  and the couplings remain perturbative up to $\Lambda\approx 10^{3}-10^4$ TeV.
Note however, that the entire low-energy theory below $F$ could have been defined without actually specifying the structure of the UV completion of the theory around $F$ and the origin of the scale $f$. 
We find it very appealing that such a simple theory perturbative up to large scales can be found. 
It is entirely possible that other UV completions with even less tuning can give the same low-energy physics around the TeV scale, for example a somewhat different anomaly free fermion matter content can be also used~\cite{othermatter}.

\begin{figure}
\bc
\includegraphics[width=0.38\textwidth]{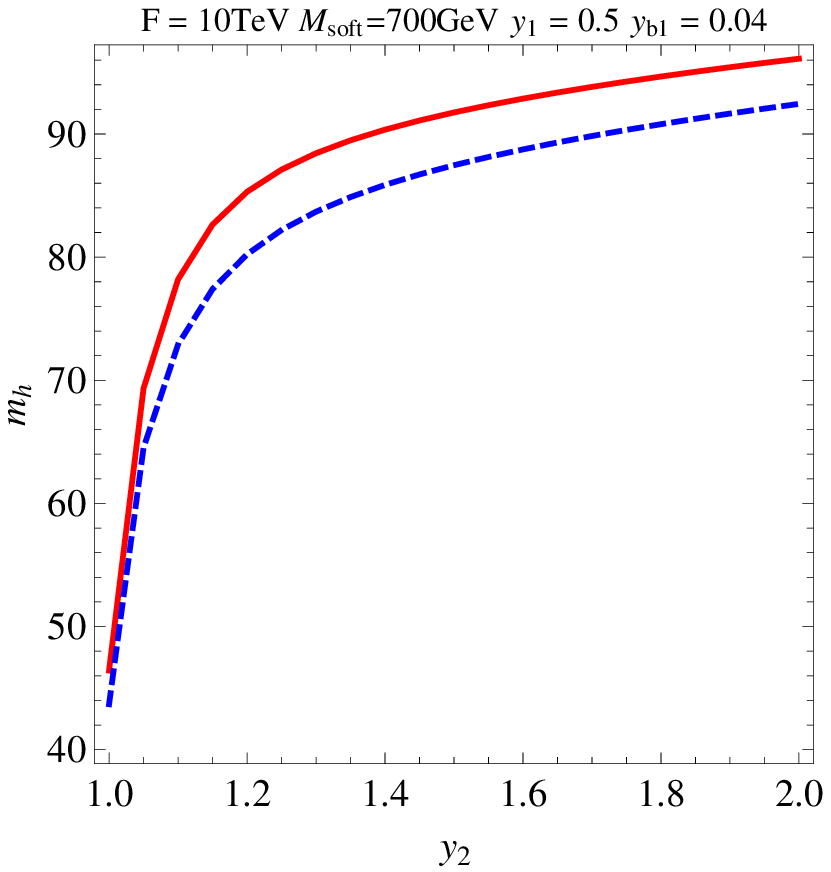}\hspace*{1.cm}
\includegraphics[width=0.38\textwidth]{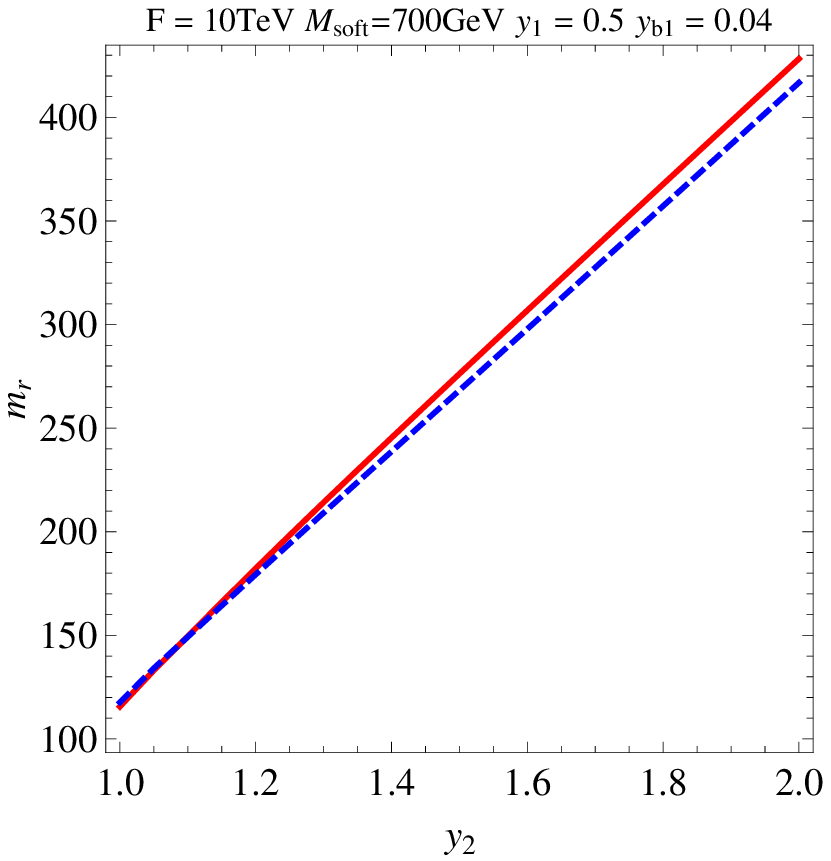}
\ec
\caption{The mass of the pGB Higgs (left panel) and the radial mode (right panel) for a sample slice of the parameter space, for $f = 350$ (dahshed blue) and $f = 400$ (solid red) GeV. This plot was obtained using the full 1-loop Coleman-Weinberg potential including the mixing between the Higgs and the radial mode. }
\label{f.higgsmass}
\end{figure}

The one-loop contributions to the pGB Higgs potential, on the other hand, are completely finite and calculable.
Electroweak symmetry breaking is triggered by negative contributions to the Higgs mass parameter from top/stop loops,
\beq
\Delta m^2 \approx - {3 m_t^2 \over 8 \pi^2 v_{EW}^2}
\left [
 M_{T}^2  \log{\msusy^2 +  M_{T}^2 \over  M_{T}^2}
+ \msusy^2 \log{\msusy^2 +  M_{T}^2  \over \msusy^2}
\right ], 
\eeq
while the contributions from the bottom sector are down by $m_b^2/m_t^2 \ll 1$.
There are also one-loop contributions to the pGB Higgs quartic, and the Higgs boson mass is
\begin{eqnarray}
m_h^2 &=& \left (1 - {v_{EW}^2 \over f^2} \right )
\left\{ m_Z^2 \cos^2 (2\beta)   +
{3 m_t^4 \over 4 \pi^2 v_{EW}^2}
\left [ \log \left (\msusy^2 M_{T}^2 \over m_t^2(\msusy^2 + M_{T}^2)\right )\right. \right. \nonumber \\
&& \left. \left.- 2 {\msusy^2 \over M_{T}^2} \log \left ( \msusy^2 + M_{T}^2  \over \msusy^2  \right )
\right ]  \right \}. 
\end{eqnarray}
Note that the tree-level Higgs boson mass is suppressed with respect to $m_Z$ by the factor of $\cos(\ti v/f)$ which is of order $0.8-0.9$ for the interesting range of $f$.
The one-loop corrections lift the Higgs boson mass above the tree-level value, but for natural values of the heavy top and soft mass they cannot add much more than 10 GeV.
As a consequence, the Higgs mass typically ends up in the 80-100 GeV range. While a complete 1-loop analysis of the full spectrum is beyond the scope of this paper (for instance we have not included the heavy Higgs scalars and the second radial mode $r_d$), we have numerically calculated the 1-loop effective Coleman-Weinberg potential for the light Higgs $h$ and the radial mode $r$ due to the top-stop and bottom-sbottom loops. We have minimized this potential in the presence of the tree-level SU(2)$\times$U(1) D-terms explicitely breaking the global symmetry and also soft breaking scalar masses. The resulting Higgs and radial masses for a typical choice of parameters is displayed in Fig.~\ref{f.higgsmass}.
The fine tuning in the doublet Higgs potential (which is usually the main source of fine tuning in the MSSM) defined as \begin{equation}
FT_2= \frac{m_h^2/2}{|\Delta m^2|}
\end{equation}
is typically completely absent in that Higgs mass range. 
In summary, in absence of additional theoretical structures such as additional gauge singlet superfields or U(1) D-terms, double protection combined with naturalness predict that the Higgs boson mass should be kinematically accessible at LEP energies.

\subsection*{$\eta$ mass}

\begin{figure}[h]
\bc
\includegraphics[width=0.38\textwidth]{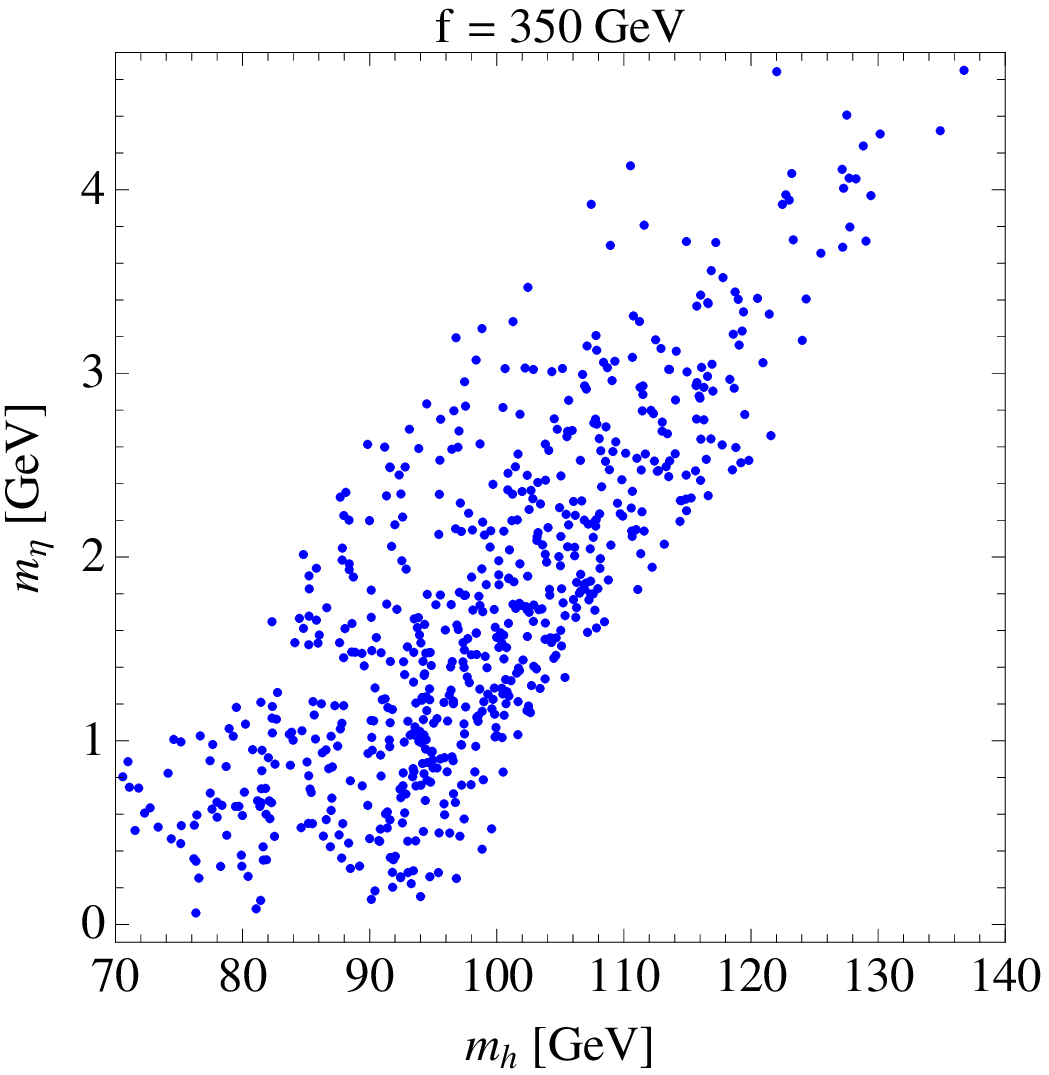}\hspace*{1.cm}
\includegraphics[width=0.38\textwidth]{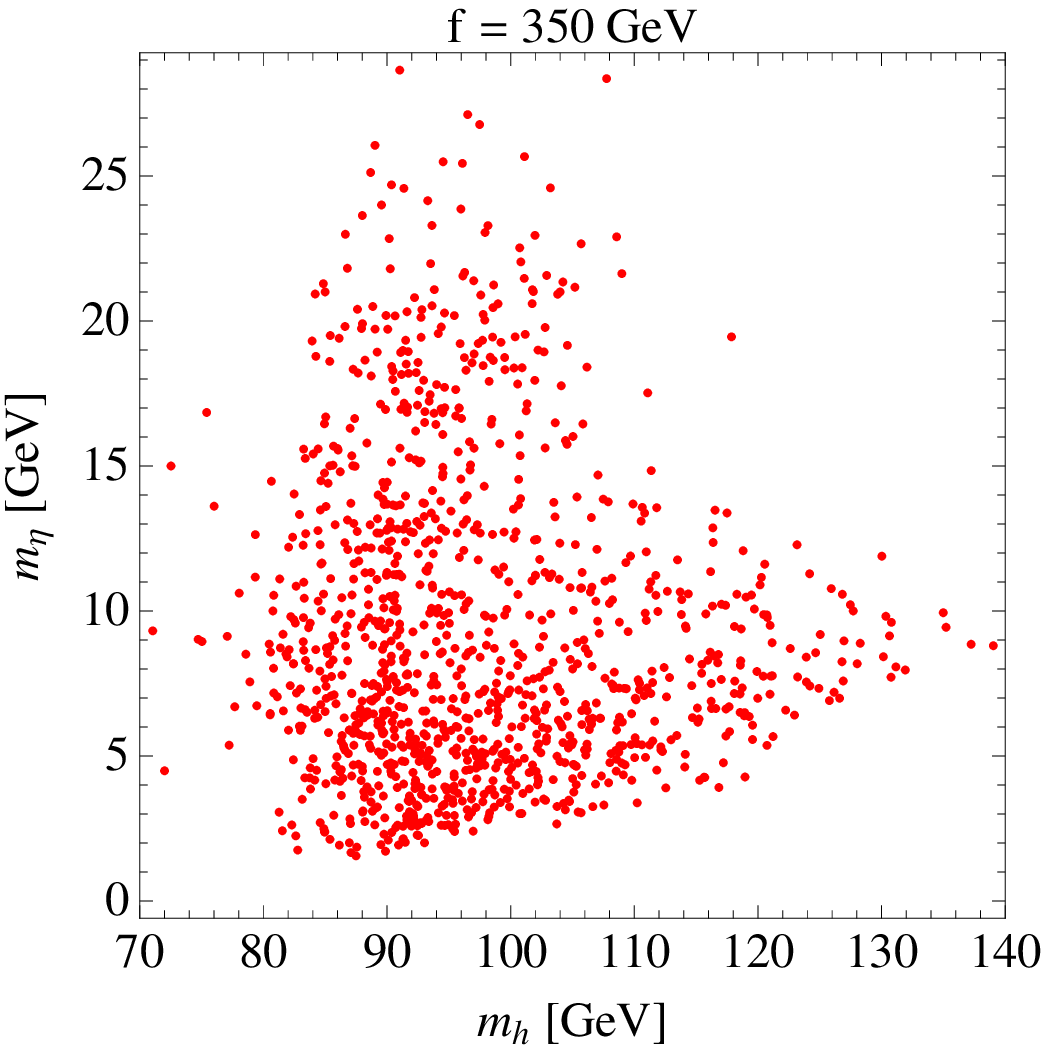}
\ec
\caption{A scan of the parameter space for the achievable $\eta$ and Higgs masses. In the left panel we show
the Higgs mass with $\tilde{y}_{b1,b2}=0$, and see that for the interesting range of Higgs masses $m_{\eta}<3$ GeV.
In the right we varied $0.001 <\tilde{y}_{b1} < 0.002$. We can see that with this non-collective Yukawa one can easily get
$m_{\eta}$ in the 5--10 GeV range.  We have fixed $f=350, F= \sqrt{2} \cdot 10^4, \Lambda=10^7$ GeV for both plots, and scanned the remaining parameters in the regions $0.02<y_1<0.3, 1<y_2<2.4, 0.02<y_{b1}<0.12$ and $300<M_{soft}<1000$ GeV.}
\label{f.etavshiggs}
\end{figure}

Fortunately, double protection in its simplest version also predicts the existence of the singlet pGB  $\eta$  via which the Higgs can cascade decay, thus avoiding discovery at LEP.
The same structure that ensures double protection of the pGB Higgs potential also ensures that the singlet mass is much smaller than the Higgs boson mass.
Indeed, in the limit of collective symmetry breaking and the gauge symmetry breaking scale $F$ going to infinity, the singlet mass vanishes.
In that limit, $\eta$ is fully embedded in the third component of the triplets $\cH_{u,d}$ and one can easily see that all non-derivative couplings of $\eta$ to the third generation  can be removed by rotations of the  top and bottom quarks by the phase factors
$e^{\pm i \ti \eta/\sqrt{2}f}$. 
After these rotations, $\eta$ has only derivative couplings to the fermions,
 in other words, $\eta$ is a genuine Goldstone boson rather than a pGB.
However, for a finite $F$, the singlet has $f/F$ suppressed components in $\Phi_{u,d}$ and the Yukawa couplings can no longer be rotated away.
The third generation quark loops then generate the operators like $|\cH_u \Phi_d|^2$, which lead to $\eta$ acquiring a mass after electroweak symmetry breaking.
The dominant contribution to the singlet  mass arises from loops of the bottom quark and its symmetry partners, and is given by (for large $\tan \beta$)
\beq
m_\eta^2 \approx
{3 v_{EW}^2 y_2^2 \over 8 \pi^2 } {\msusy^2 \over F^2}\left [
 \log\left( y_{b1}^2 F^2 \over  2 (M_T^2 + \msusy^2)\right)
-  {M_T^2 \over \msusy^2} \log\left(M_T^2 + \msusy^2 \over M_T^2\right)  + 1  \right ].
\eeq
For $F \sim 10$ TeV and $y_{b1} \sim 0.1$ this leads to $\eta$ in the 1--3 GeV range.
The top loop contribution is subleading because it is not enhanced by $\log F$.
As we will discuss soon, the above contribution is not enough to satisfy the existing LEP bounds if the Higgs mass is in the 78--86 GeV window, in  which case  $6 \gev  \simlt m_\eta < 9.2$ GeV is required.
In such a case, we can always invoke a small addition of non-collective couplings that enhance the one-loop contributions to the $\eta$ mass without spoiling naturalness.
For example, we can add non-collective Yukawa couplings in the bottom sector,
\begin{equation}
\tilde{y}_{b1} \mathcal{H}_d Q b_c^1 +\tilde{y}_{b2} \Phi_d Q b_c^2, 
\label{noncollective}
\end{equation}
with $\ti y_{b} \sim 10^{-3}$.
This leads to additional one-loop contributions to the $\eta$ mass approximately given by
\begin{equation}
m_{\eta}^2 = \cos \beta \frac{N_c}{4\pi} \frac{F}{f} (y_{b1}\tilde{y}_{b1}+y_{b2}\tilde{y}_{b2}) M_{soft}^2 \log \frac{\Lambda}{F}.
\end{equation}
In Fig.~\ref{f.etavshiggs} we show a scatterplot of the Higgs and $\eta$ masses by varying the input parameters both without and with the non-collective bottom Yukawas.

\subsection*{Parameter scans}

\begin{figure}
\bc
\includegraphics[width=0.38\textwidth]{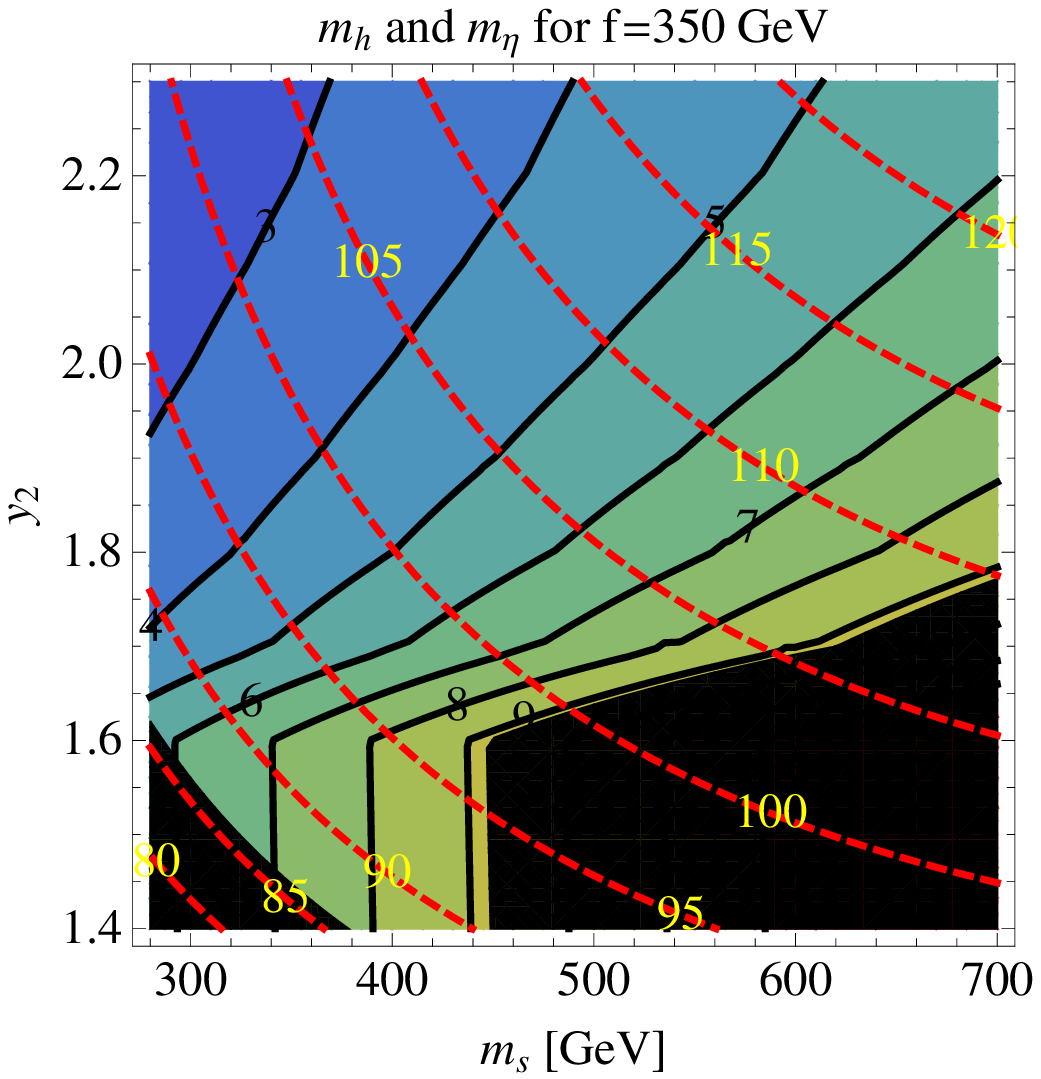}\hspace*{1.cm}
\includegraphics[width=0.38\textwidth]{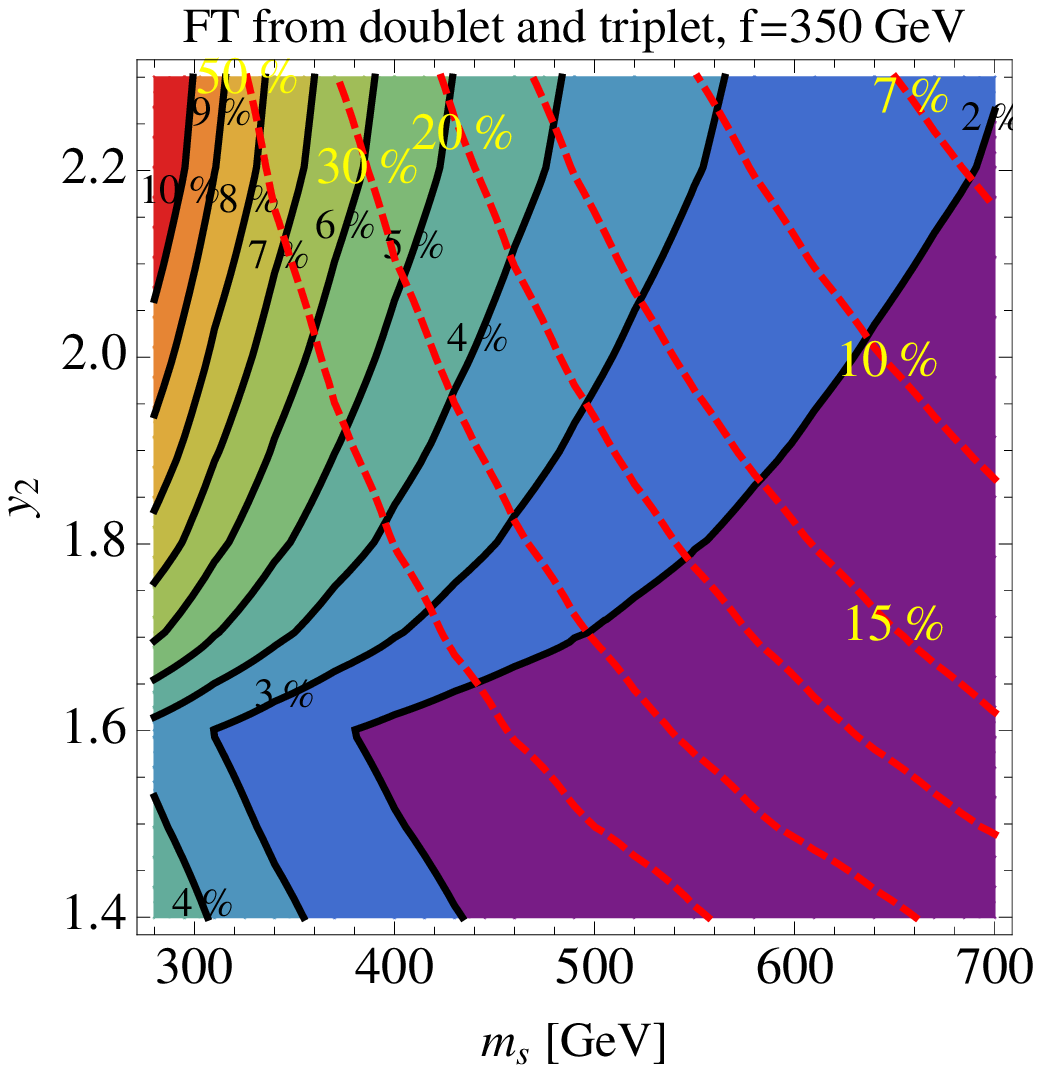}
\ec
\caption{On the left the contours of the Higgs mass (dashed red line) and the $\eta$ mass (solid black lines) as function of the universal soft breaking mass $M_{soft}$ and the top Yukawa $y_2$. On the right, the necessary fine tunings $FT_3$ (solid black) and $FT_2$ (dashed red) in percent. The kink in the contour lines at $y_2\approx 1.64$ appears because the cut-off for larger values is determined by the Landau pole of $y_2$.
These plots are based on the full numerical 1-loop Coleman-Weinberg potential,  with $f=350$ GeV, $y_1= 0.29, y_{b1} = 0.1, y_{b2} = 0, \tilde{y}_{b1} = 0.001,  \tan\beta = 10, F = \sqrt{2}\cdot 10^4$ and $\tilde{y}_{b2}= 0$.
The region in the lower left is excluded by the LEP $\xi^2 {\mathrm{BR}}(h \to b \ov b)$ bound and in the lower right because $m_{\eta}>2 m_b$.}
\label{f.350GeV}
\end{figure}

\begin{figure}
\bc
\includegraphics[width=0.38\textwidth]{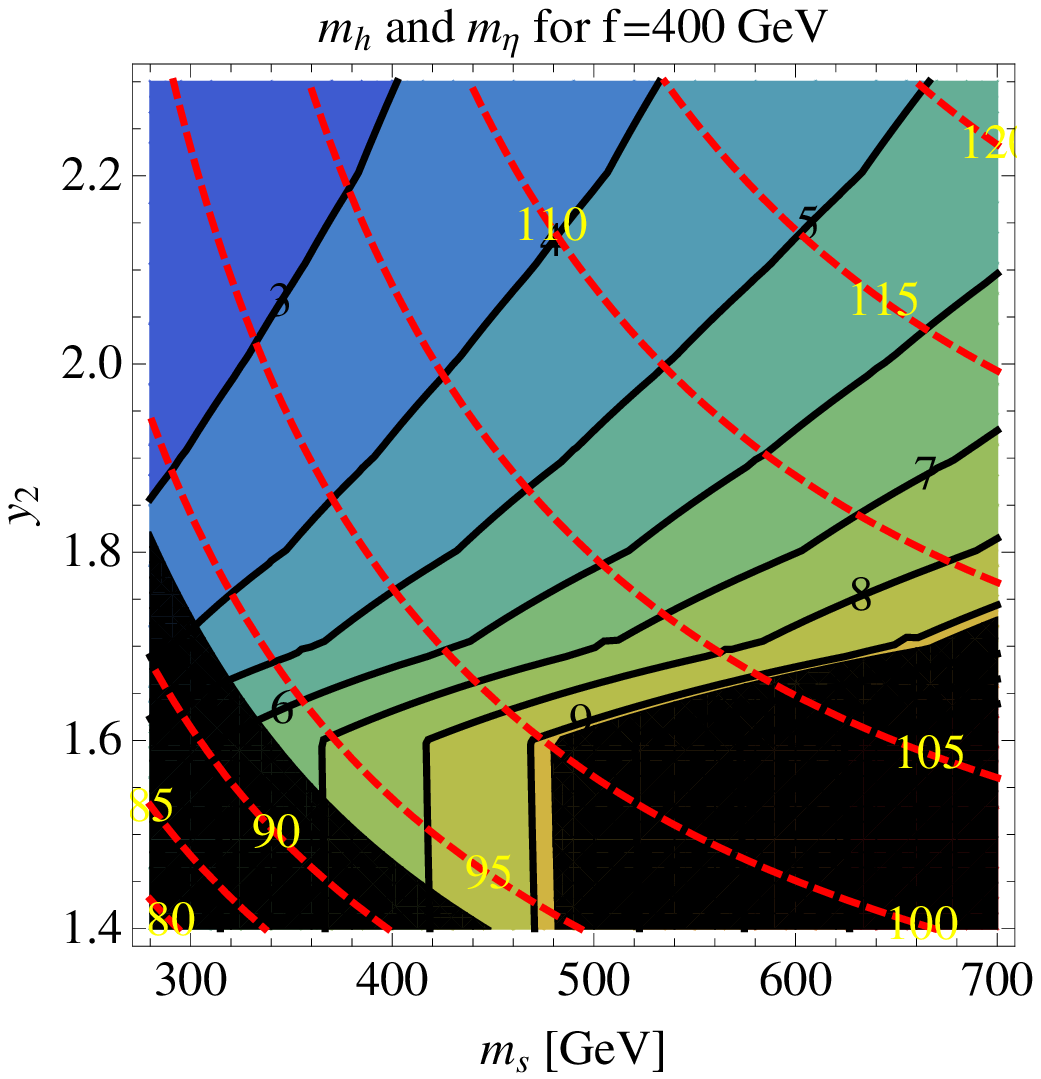}\hspace*{1.cm}
\includegraphics[width=0.38\textwidth]{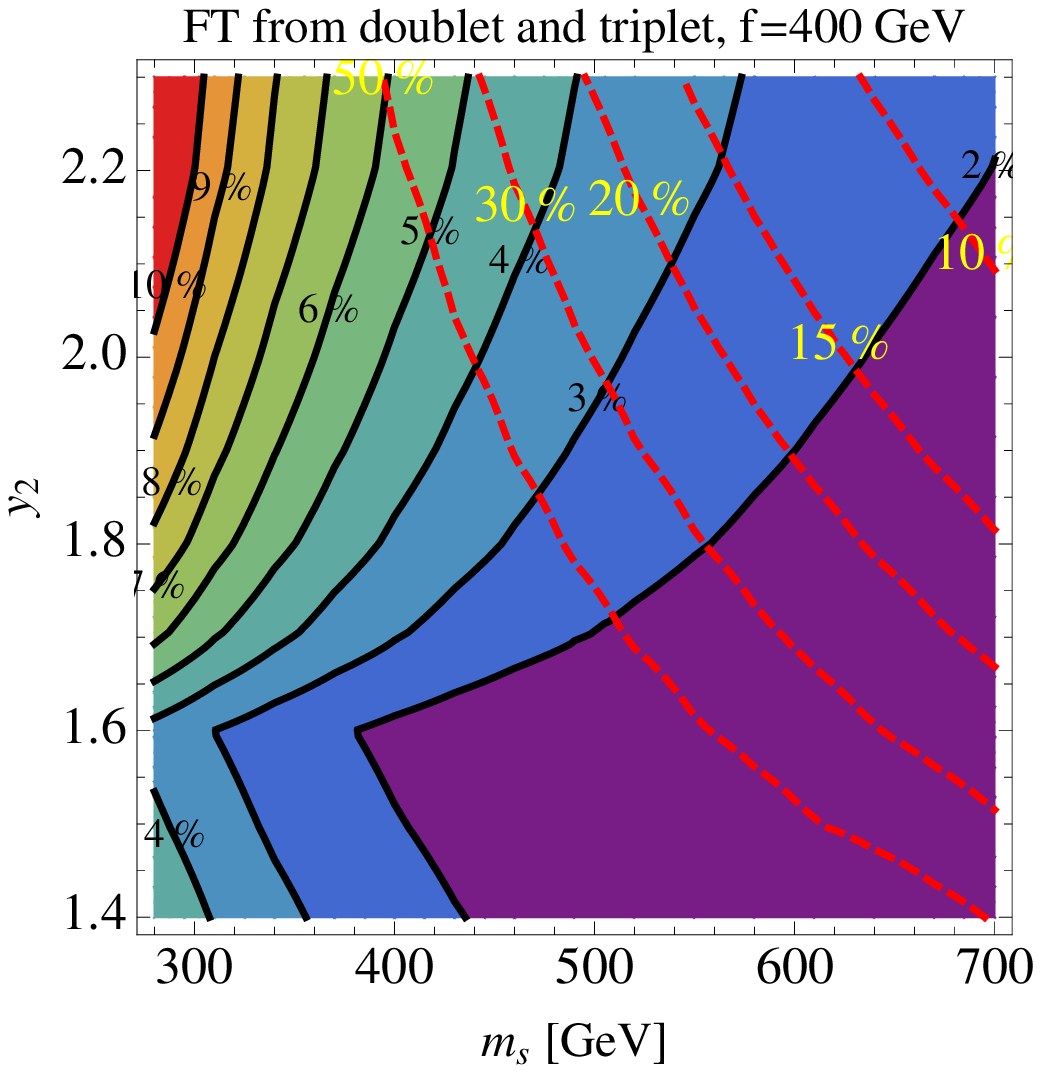}
\ec
\caption{The same as in~\ref{f.350GeV} for $f=400$ GeV.}
\label{f.400GeV}
\end{figure}

To illustrate the parameter space achievable in this model we have prepared two sets of contour plots for $f=350$ (Fig.~\ref{f.350GeV}) and $400$ GeV (Fig.~\ref{f.400GeV}), where we show both the Higgs and $\eta$ masses and the fine tunings $FT_{3,2}$. While these scans are not exhaustive, we can see that large regions of the parameter space are open with Higgs masses in the $85-115$ GeV range and $\eta$ masses between 3 and 9 GeV. 
The usual MSSM fine tuning $FT_2$ is mostly negligible, while the UV completion dependent fine tuning $FT_3$ varies in the 3-10\% range.
For example,  for $f=350$ GeV
and a relatively  small top Yukawa coupling $y_{2}\approx 1.64$, 
the theory would stays perturbative up to $\Lambda\approx 10^{8}$ TeV  (which is also the Landau pole for the strong coupling $g_{3}$ in the presence of one family of the vectorial states $(V,V_{c})$), while the fine tuning $FT_3$ is about $5$\% for $M_{soft} \sim 300$ GeV and a $90$ GeV Higgs mass and $6$ GeV $\eta$ mass.

\subsection*{$\eta$ decays: hiding the Higgs at LEP (and the LHC)}

The last part of the plan to allow the Higgs to escape detection at LEP is to ensure that $\eta$ decays via a channel that is not well constrained by LEP.
This is possible only if the singlet is lighter than twice the b quark mass; if kinematically allowed, the $\eta \to b \bar b $ decay channel always dominates which excludes the Higgs lighter than 110 GeV.
In the NMMSM context, the dominant decay below $2b$ threshold is the decay into tau leptons.
In our model this is not the case.
The reason is that the pGB singlet is embedded in the third component of the $\cH_{u,d}$ triplets and it can couple to the SM quarks and leptons only via their mixing with the heavy fermionic states.
For $\tau$, that mixing angle is suppressed by $m_\tau^2/M_\tau^2$, where the heavy tau mass should be larger than few $\times 100$ GeV.
More precisely, the coupling of the pGB singlet is
\beq
\label{e.etatau}  
i \ti y_\tau (\bar \tau \gamma_5 \tau) \eta   \qquad
\ti y_\tau \simeq {m_\tau^3 f \over \sqrt{2} M_\tau^2 v_{EW}^2} 
\eeq
where $M_\tau \approx y_{\tau 1} F/\sqrt{2}$.  The corresponding decay width is
\beq
\Gamma_{\eta\to\tau\tau} \approx  {1 \over 16 \pi} \sqrt{1 - 4 m_\tau^2/m_\eta^2}
{m_\eta m_\tau^6 f^2 \over  v_{EW}^4 M_\tau^4} . 
\eeq
It depends on the sixth power of the tau mass and for this reason it is much more suppressed than in the NMSSM models.
For typical parameters, $f \sim 350$ GeV and $M_\tau \sim 200$ GeV, the width into tau is in the $10^{-14} - 10^{-13}$ GeV range  corresponding to order millimeter decay length.

\begin{figure}
\bc
\includegraphics[width=0.25\textwidth]{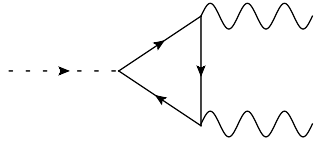}
\ec
\caption{The diagram for one loop $\eta$ decay into gluons or photons.}
\label{f.egg}
\end{figure}

Because of the suppression of the $\eta\tau\bar{\tau}$ coupling, the pGB singlet decays dominantly into two gluons, via the loop diagram  in \fref{egg}  with bottom and top and their fermionic partners running in the loop (the scalar partners do not contribute to this decay amplitude).
Quite generally, starting with the coupling $i \ti y_\psi \eta (\ov \psi \gamma_5 \psi)$ to light or heavy fermions, one-loop effects generate the effective coupling  \cite{Spira}
\beq
\label{e.kappa}
\kappa^g \eta \eps^{\mu\nu\rho\sigma} G_{\mu \nu}^a  G_{\rho \sigma}^a \ ,
\qquad
\kappa^g = {g^2 \over 32 \pi^2} \sum_{\psi} {\ti y_\psi \over  m_\psi}  c_2(\psi) \tau_\psi f(\tau_\psi)
\eeq
where
\beq
\tau_\psi = 4 m_\psi^2/m_\eta^2
\qquad
 f(\tau) = \left \{ \ba{cc}
\arcsin^2[\tau^{-1/2}] & \tau \geq 1
\\
-{1 \over 4} \left (\log[(1 + \sqrt{1 - \tau})/(1 - \sqrt{1 - \tau})] - i \pi \right)^2
& \tau < 1
\ea \right .
\eeq
Furthermore, $g = g_s(m_\eta)$ -  the color SU(3) coupling at the scale of the singlet mass and $c_2$ is the Dynkin index of the quark representation which  is equal to $1/2$ for the fundamental representation.
There is an analogous coupling $\kappa^\gamma$ to the photon field strengths with $g \to g_{em}$ and  $c_2 \to N_c Q_\psi^2$, $N_c = 3$.
The decay width into two gluons and two photons is given by
\beq
\Gamma_{\eta \to gg} = (N_c^2 - 1){|\kappa^g|^2 \over \pi} m_\eta^3,
\qquad
\Gamma_{\eta \to \gamma\gamma} = {|\kappa^\gamma|^2 \over \pi} m_\eta^3.
\eeq
The pGB singlet has the largest coupling to the bottom and the top quarks,
\beq
\ti y_t \simeq  {m_t^3 f \over \sqrt{2} v_{EW}^2 \mu_V^2},
\qquad
\ti y_b \simeq  {m_b m_t^2 f \over \sqrt{2} v_{EW}^2 \mu_V^2}.
\eeq
Since $\kappa^g \sim \ti y_\psi/m_\psi$ for $m_\eta \ll 2 m_\psi$ one would expect that the top and bottom loops dominate the decay amplitude and give roughly the same contribution.
This is not quite correct.
In the model at hand there is a sum rule $\sum \ti y_\psi/m_\psi \approx \co(1/F^2)$ separately in the bottom, top, and tau sectors.
This sum rule is the consequence of the fact that $\eta$, at the leading order in $1/F$, couples to a gauge symmetry current that is necessarily anomaly free (one can see in the parametrization of \eref{pgbe} that $\eta$ couples to a combination of the $SU(3)$ $T_8$ generator  and  $U(1)_X$).
Thus, the operator $\eta G \ti G$ cannot be generated by integrating out fermions;
the lowest allowed operator is $\Box \eta G \ti G$.
As a consequence, the amplitude is proportional to $\sum \ti y_\psi m_\eta^2/m_\psi^3$, which is non-vanishing and largely dominated by the SM bottom quark contribution.
One finds
\beq
\kappa^g  \simeq {1 \over 12 \sqrt 2} {g_s^2(m_\eta) \over 64 \pi^2} {m_\eta^2 \over m_b^2}
{m_t^2 f  \over \mu_V^2 v_{EW}^2}. 
\eeq
For the photons, one should replace  $g_s^2/2 \to N_c g_{em}^2 Q_b^2$.
The bottom-loop domination of the decay amplitude has a practical consequence that the photonic branching ratio is more suppressed than in the SM because $|Q_b| = |Q_t|/2$.
At the end of the day the ${\mathrm{BR}}(\eta \to \gamma \gamma)$ is of order $10^{-4}$.
While this is not of much relevance for the LEP searches,
the additional suppression will make the LHC Higgs search even more difficult if possible at all.

\begin{figure}
\bc
\includegraphics[width=0.38\textwidth]{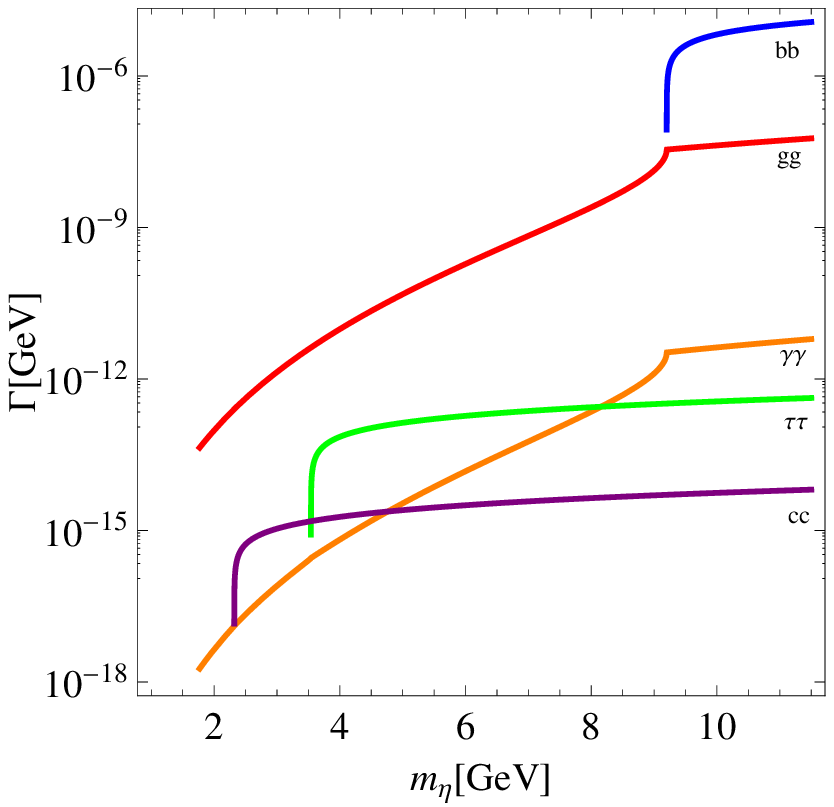}\hspace*{1.cm}
\includegraphics[width=0.38\textwidth]{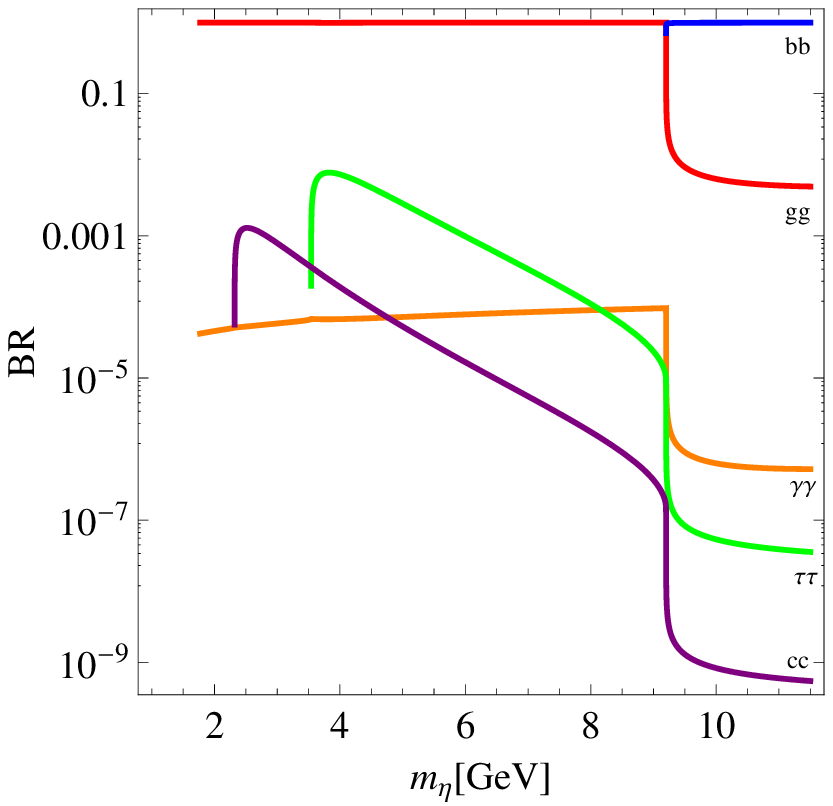}
\ec
\caption{The partial widths
and the branching ratios
of the pGB singlet $\eta$ for decays into $gg$, $\gamma \gamma$, $bb$, $\tau\tau$ and $cc$.
The parameters are $f = 350$ GeV, $\mu_V = 500$ GeV, $M_{c} = 400 GeV$, $M_\tau = 200$ GeV. }
\label{f.etabr}
\end{figure}

The partial widths and the branching ratios of $\eta$ are plotted in \fref{etabr}.
One can see that below the 2b threshold the dominant decay channel of the Higgs is the four-gluon cascade decay.
The branching ratio for $h \to gg\gamma\gamma$ is of order $10^{-4}$.
Discovering the Higgs decaying almost exclusively to 4 gluons with such a small branching ratio into photons might be impossible at the LHC \cite{CFW}.
Fermionic decay channels are also hugely suppressed, for example the branching ratio for  $h \to gg\tau^+ \tau^-$ is in the range $10^{-5} - 10^{-3}$.
This fermiophobic feature of $\eta$ implies that the recent D0 searches of  $h \to \tau \tau \mu \mu$ and $h \to 4\mu$ \cite{D0search,LW},
as well as the BaBar and CLEO studies of $\upsilon$ decays \cite{BaBar,DGM} do not constrain the parameter space of our model.
Furthermore, we estimate the branching ratio for $\upsilon \to \gamma + \eta$ to be of order $10^{-5}$, which is safely below the CLEO limit of $10^{-4}$ \cite{rosner}.

The best limits on the Higgs mass in our scenario follow from the analyses published by the OPAL collaboration.
Higgs masses smaller than $\sim 78$ GeV are excluded by the decay-mode independent search \cite{OPALgeneral}.
A search for $h\to 4j$ via a pseudoscalar has been performed for the $78-86$ GeV  Higgs mass window in \cite{OPAL4g}.
In our model, the Higgs branching fraction into 4 jets is of order 80\%, and moreover there is the suppression of $1-v_{EW}^2/f^2$ of the Higgs production cross section.
This implies that already in the existing OPAL search the $\eta$ mass in the range $6 \gev \simlt m_\eta \simlt 9.2 \gev$ is allowed, which can be achieved in our model in the presence of small non-collective Yukawa couplings.
The OPAL bound on the $\eta$-mass can be understood qualitatively quite simply.\footnote{We thank Paddy Fox for explaining this to us.} When $m_{\eta}$ is very small, the etas will be highly boosted and both pairs of gluon jets will be collimated. The angle between the two gluon jets is of order $4 m_{\eta}/m_h$, which for very small $m_{\eta}$ is too small for the four jets to be independently resolved, in which case the very restrictive $h\to 2j$ exclusion limit will apply. For Higgs masses above $86$ GeV our scenario is not constrained by any existing experimental analysis.
Extrapolating the bound from \cite{OPAL4g} one expects that the bound on $m_{\eta}$ will weaken with increasing Higgs mass, and will dip below few GeV when $m_h \simgt 90$ GeV.

\subsection*{Electroweak precision, unitarity and the radial mode}

In our model,  the pGB Higgs coupling to W and Z is suppressed by $\cos (\ti v/f)$ with respect to the SM value.
Because of that,  the cancellation of logarithmic divergences in the gauge boson self-energies is complete only after taking into account
the radial modes $r_{u,d}$ associated with the oscillations of the VEVs of $\cH_{u,d}$.
For large or moderate $\tan \beta$ it is $r = r_u$ that has sizable couplings to W and Z: that coupling is suppressed by $\sin (\ti v/f) = {v_{EW}/f} \sim 1/2$ with respect to the SM Higgs coupling.
The fact that there are two Higgs-like particle coupled to the SM gauge bosons affects electroweak precision observables.
 As pointed out in~\cite{BBRV}, since the electroweak $S$ and $T$  parameters depend logarithmically on the Higgs mass, one can estimate the effect on $S$ and $T$ by defining an effective Higgs mass $m_{EWPT}$ which at large $\tan\beta$ is given by
 \begin{equation}
 m_{EWPT}=m_{h}\left( \frac{m_{r}}{m_{h}}\right)^{v_{EW}^{2}/f^{2}}, 
 \end{equation}
where $m_{r}^{2}\approx4\lambda_{\cH_{u}} f^{2}$ is the mass of the radial mode $r$.
This is because the SM $\log m_h^2/\Lambda^2$ dependence is expected to be replaced everywhere by $ \cos^2 (\ti v/f) \log m_h^2/\Lambda^2 +\sin^2 (\ti v/f) \log m_r^2/\Lambda^2$. For typical range of
parameters $m_{r}\sim 300-400$ GeV the  corresponding  $m_{EWPT}$ is in the safe $110-135$ GeV  range, so one expects the oblique corrections to be within the experimentally preferred region. Another potential electroweak precision correction
 is the tree-level shift in the $Zb\bar{b}$ vertex, which is due to the mixing of the physical left-handed bottom $b$ with the states $\hat{b}^Q$ (and also the right handed $b$ with $b_c^V$). An explicit calculation shows that the shift of the left handed $Zb\bar{b}$ vertex is of the order
\begin{equation}
\frac{\delta g}{g} \propto \cos^2 \beta \frac{\mu_V^2 v_{EW}^2}{m_{B1}^4 m_{B2}^2}\left[ F (y_{b1} \tilde{y}_{b1}+y_{b2} \tilde{y}_{b2})+ f \cos\beta (y_{b2}^2+\tilde{y}_{b1}^2) \right]^2,
\end{equation}
where $m_{B1,B2}$ are the heavy bottom masses in the multi-hundred GeV range. We can see that the shift is suppressed by $\cos^2 \beta \sim 0.01$, and has large additional mass suppression factors reducing it well below the experimental bound of a few per mille.
There are also one-loop corrections to the $Zb\bar{b}$ vertex, and to the $S$ and $T$ parameters as well, coming for instance from the extended Yukawa sector. While they deserve future investigations, a full one-loop analysis is beyond the scope of the present paper. 

The radial mode also takes part in unitarizing WW scattering.
As usual in pseudo-Goldstone Higgs theories, the $\cos (\ti v/f)$ suppression of the pGB Higgs coupling to the SM gauge boson leads to an imperfect cancellation of the contributions that grow with $E^2$ in the longitudinal gauge boson scattering amplitudes \cite{GGPR}.
In our case however the theory is weakly coupled as the radial mode completes the cancellation well below  the unitarity bound of $\sim 1$ TeV.
Thus we do not expect a significant rise in the $WW$ scattering amplitude as long as the mass of the radial mode is within the expected $300-400$ GeV range.

The radial mode $r$ itself might actually be more easily visible at the LHC than the Higgs $h$.
Since the radial mode has (suppressed) Higgs-like couplings and mass in the $300-400$ GeV range, the standard heavy Higgs decay to $WW$ and $ZZ$ is allowed.
In particular, $r\to ZZ\to 4l$ mode should be open and easily observable at the LHC despite the fact that the production rate is suppressed by $v_{EW}^2/f^2$ as compared to the SM Higgs.
This may lead to the peculiar situation where the LHC discovers a ``wrong" Higgs $r$ which  by itself cannot account for either the electroweak precision fit or the unitarization of the WW scattering, while the signal of the true Higgs is swamped by the QCD background.
Apart from the production cross section and couplings to gauge bosons, another phenomenological difference with the standard Higgs would be a suppression of the $h \to tt$ branching ratio: the symmetry protection of the Higgs potential typically results in a reduced Yukawa coupling with the top for both the pGB Higgs and the radial mode.

\subsection*{Conclusions}

We have presented a simple complete model where the traditional tuning in the MSSM Higgs potential is absent due to the pseudo-Goldstone nature of the Higgs. 
No additional sectors have to be included to raise the Higgs quartic because the Higgs dominantly decays to the fifth pseudo-Goldstone boson $\eta$, which can hide the Higgs at LEP. 
A tree-level coupling responsible for the $h\to 2\eta$ is automatically present, and the mass of $\eta$ is naturally in the few to 10 GeV range.
The leading Higgs decay turns out to be $h\to 2\eta \to 4 j$, which is not excluded by the existing LEP analyses for $m_h> 86 $ GeV,
and for $\eta$ masses in the 6--9.2 GeV window it is also not excluded by the dedicated OPAL analyses in the $78 < m_h < 86$ GeV window.
While this Higgs is ``buried" in the QCD background and would remain elusive at the LHC,  a heavy Higgs-like radial mode and many of the symmetry partners of the SM fermions would be readily observable.

\subsection*{Acknowledgments} We thank Jim Alexander, Roberto Contino, Paddy Fox, Bob McElrath, Maxim Perelstein, Matt Reece, Gavin Salam, Jay Wacker, Lian-Tao Wang, Neal Weiner, Alvise Varagnolo, Itay Yavin, and Dieter Zeppenfeld for useful discussions and comments.  This research of B.B. and C.C. has been supported in part by the NSF grant PHY-0757868. C.C. was also supported in part by a U.S.-Israeli BSF grant.
The work of A.F. was supported in part by the Department of Energy grant DE-FG02-96ER40949.


\end{document}